\def\sgline{\noalign{\vskip 0.15truecm\hrule\vskip 0.15truecm}}
\def\piz{\pi^0}
\def\calB{\mbox{${\cal B}$}}
\newcommand{\gaga}{{\gamma\gamma}}
\newcommand{\etapr}{{\eta^{\prime}}}
\newcommand{\Bomegapi}{\mbox{$B^+\rightarrow\omega\pi^+$}}
\newcommand{\Bomegak}{\mbox{$B^+\rightarrow\omega K^+$}}
\newcommand{\Bomegah}{\mbox{$B^+\rightarrow\omega h^+$}}

\newcommand{\Bomegaetapr}{\mbox{$B^0\rightarrow\omega\etapr$}}
\newcommand{\Bomegaeta}{\mbox{$B^0\rightarrow\omega\eta$}}
\newcommand{\Bomegarhop}{\mbox{$B^+\rightarrow\omega \rho^+$}}
\newcommand{\Bomegarhoz}{\mbox{$B^0\rightarrow\omega \rho^0$}}
\newcommand{\Bomegakz}{\mbox{$B^0\rightarrow\omega K^{0}$}}
\newcommand{\Bomegapiz}{\mbox{$B^0\rightarrow\omega \pi^{0}$}}
\newcommand{\Bomegakstz}{\mbox{$B^0\rightarrow\omega K^{*0}$}}
\newcommand{\Bomegakstp}{\mbox{$B^+\rightarrow\omega K^{*+}$}}
\newcommand{\Bomegaomega}{\mbox{$B^0\rightarrow\omega \omega$}}
\newcommand{\Bphieta}{\mbox{$B^0\rightarrow\phi\eta$}}
\newcommand{\Bphietapr}{\mbox{$B^0\rightarrow\phi\etapr$}}

\newcommand{\omegak}{\mbox{$\omega K^+$}}
\newcommand{\omegakz}{\mbox{$\omega K^0$}}
\newcommand{\omegapi}{\mbox{$\omega\pi^+$}}
\newcommand{\omegapiz}{\mbox{$\omega\pi^0$}}
\newcommand{\omegah}{\mbox{$\omega h^+$}}

\newcommand{\omegaetaprd}{\mbox{$\omega\etapr_{\eta\pi\pi}$}}
\newcommand{\omegaetaprrg}{\mbox{$\omega\etapr_{\rho\gamma}$}}

\newcommand{\omegaetagg}{\mbox{$\omega\eta_{\gaga}$}}
\newcommand{\omegaetathrp}{\mbox{$\omega\eta_{3\pi}$}}

\newcommand{\omegakstzd}{\mbox{$\omega K^{*0}_{K^+\pi^-}$}}

\newcommand{\omegakstpd}{\mbox{$\omega K^{*+}_{K^+\piz}$}}
\newcommand{\omegakstpkz}{\mbox{$\omega K^{*+}_{K^0\pi^+}$}}
\newcommand{\omegarhoz}{\mbox{$\omega \rho^0$}}
\newcommand{\omegarhop}{\mbox{$\omega \rho^+$}}
\newcommand{\omegaomega}{\mbox{$\omega\omega$}}

\newcommand{\Bphik}{\mbox{$B^+\ra\phi K^+$}}
\newcommand{\Bphikz}{\mbox{$B^0\ra\phi K^0$}}
\newcommand{\Bphipi}{\mbox{$B^+\ra\phi\pi^+$}}
\newcommand{\Bphipiz}{\mbox{$B^0\ra\phi\pi^0$}}
\newcommand{\Bphikstz}{\mbox{$B^0\ra\phi K^{*0}$}}
\newcommand{\Bphikstp}{\mbox{$B^+\ra\phi K^{*+}$}}
\newcommand{\Bphikst}{\mbox{$B\ra\phi K^*$}}
\newcommand{\Bphirhoz}{\mbox{$B^0\ra\phi \rho^0$}}
\newcommand{\Bphirhop}{\mbox{$B^+\ra\phi \rho^+$}}
\newcommand{\Bphiomega}{\mbox{$B^0\ra\phi\omega$}}
\newcommand{\Bphiphi}{\mbox{$B^0\ra\phi\phi$}}

\newcommand{\phik}{\mbox{$\phi K^+$}}
\newcommand{\phikz}{\mbox{$\phi K^0$}}
\newcommand{\phipi}{\mbox{$\phi\pi^+$}}
\newcommand{\phipiz}{\mbox{$\phi\pi^0$}}

\newcommand{\phietaprd}{\mbox{$\phi\etapr_{\eta\pi\pi}$}}
\newcommand{\phietaprrg}{\mbox{$\phi\etapr_{\rho\gamma}$}}

\newcommand{\phietagg}{\mbox{$\phi\eta_{\gaga}$}}
\newcommand{\phietathrp}{\mbox{$\phi\eta_{3\pi}$}}

\newcommand{\phikstzd}{\mbox{$\phi K^{*0}_{K^+\pi^-}$}}
\newcommand{\phikstzkz}{\mbox{$\phi K^{*0}_{K^0\piz}$}}

\newcommand{\phikstpd}{\mbox{$\phi K^{*+}_{K^+\piz}$}}
\newcommand{\phikstpkz}{\mbox{$\phi K^{*+}_{K^0\pi^+}$}}

\newcommand{\phirhoz}{\mbox{$\phi \rho^0$}}
\newcommand{\phirhop}{\mbox{$\phi \rho^+$}}
\newcommand{\phiomega}{\mbox{$\phi\omega$}}
\newcommand{\phiphi}{\mbox{$\phi\phi$}}

\newcommand{\ra}{\rightarrow}

\newcommand{\psfile}[3][]{ 
  \begin{center}
    \setlength{\epsfxsize}{#3\linewidth}\leavevmode
    \def\noOpt{}\def\testit{#1}\ifx\testit\noOpt%
      \epsfbox{#2}%
    \else%
      \epsfbox[#1]{#2}%
    \fi
  \end{center} 
}

\documentstyle[aps,prl,preprint,floats,epsfig]{revtex}

\begin{document}

\preprint{\tighten\vbox{\hbox{\hfil CLNS 97/1537}
                        \hbox{\hfil CLEO 97-32}
}}

\title{Observation of \Bomegak \\
 and Search for Related $B$ Decay Modes.}

\author{CLEO Collaboration}
\date{\today}

\maketitle
\tighten

%
%

\begin{abstract}

 We have searched for two-body charmless decays of $B$ mesons to purely
 hadronic exclusive final states including $\omega$ or $\phi$ mesons using
 data collected with the CLEO II detector. With this sample of $6.6\times
 10^6$ $B$ mesons we observe a signal for the $\omega K^+$ final state, and
 measure a branching fraction of ${\cal B}(B^+ \rightarrow \omega K^+) =
 (1.5^{+0.7}_{-0.6}\pm0.2)\times 10^{-5}$. We also observe some evidence for
 the $\phi K^*$ final state, and upper limits are given for 22 other decay
 modes. These results provide the opportunity for studies of
 theoretical models and physical parameters.

\end{abstract}

\newpage

{
\renewcommand{\thefootnote}{\fnsymbol{footnote}}

\begin{center}
T.~Bergfeld,$^{1}$ B.~I.~Eisenstein,$^{1}$ J.~Ernst,$^{1}$
G.~E.~Gladding,$^{1}$ G.~D.~Gollin,$^{1}$ R.~M.~Hans,$^{1}$
E.~Johnson,$^{1}$ I.~Karliner,$^{1}$ M.~A.~Marsh,$^{1}$
M.~Palmer,$^{1}$
% ...
\end{center}
 
\small
\begin{center}
$^{1}${University of Illinois, Urbana-Champaign, Illinois 61801}\\
% ...
\end{center}

\setcounter{footnote}{0}
}
\newpage

%
%

\begin{figure}[htbp]
\psfile{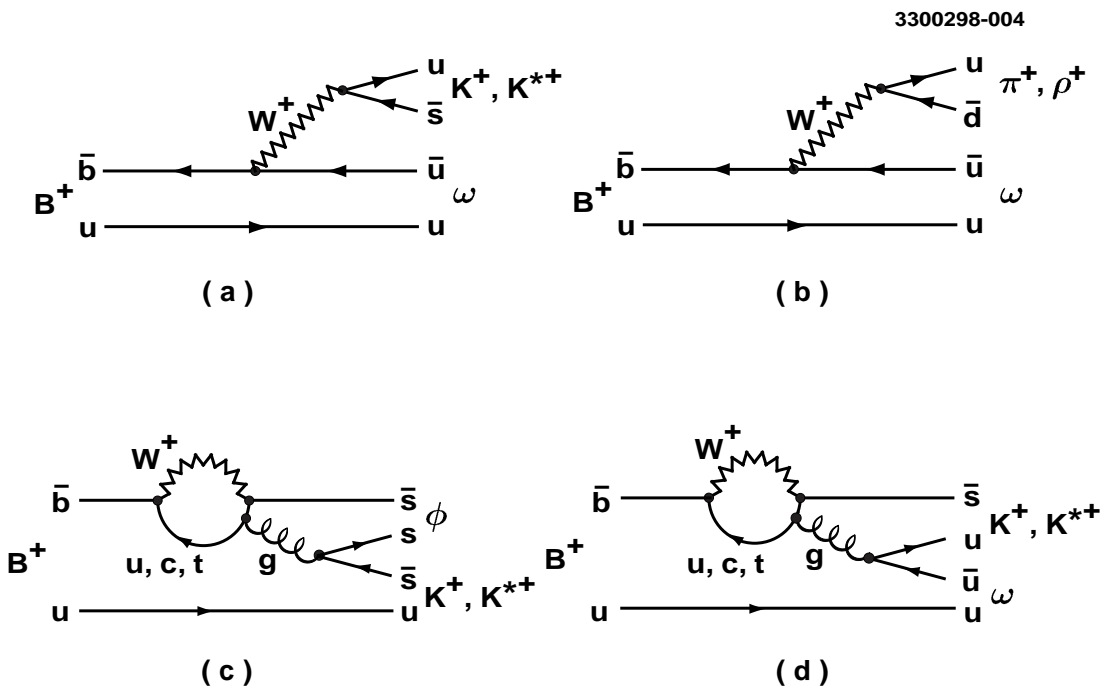}{0.9}
 \caption{\label{f:diag}%
(a) and (b) tree-level spectator and (c) and (d) penguin diagrams for some of
the decay modes investigated.
 }
\end{figure}

 In the last several years, the study of charmless non-leptonic decays of $B$
 mesons has attracted a lot of attention, primarily because of the importance
 of these processes in understanding the phenomenon of $CP$ violation. This
 interest is expected to continue as several new experimental facilities
 specifically built for $B$ meson studies begin operating within a few
 years. Purely hadronic decays of $B$ mesons are understood to proceed mainly
 through the weak decay of a $b$ quark to a lighter quark, while the light
 quark bound in the $B$ meson remains a spectator, as shown by the Feynman
 diagrams in Figure~\ref{f:diag}. The decay amplitude for ``tree-level'' $b
 \rightarrow u$ transitions (Figure~\ref{f:diag}(a) and (b)) is much smaller
 than the one for dominant $b \rightarrow c$ transitions due to the ratio
 of Cabibbo-Kobayashi-Maskawa~\cite{ckm} matrix elements $V_{ub}/V_{cb}\approx
 0.1$. Transitions to $s$ and $d$ quarks are effective flavor-changing
 neutral currents proceeding mainly by one-loop ``penguin'' amplitudes, and
 are also suppressed. Examples are shown in Figure~\ref{f:diag} (c) and (d).

 The strong interaction between particles in the final state makes theoretical
 predictions difficult. The use of effective hamiltonians, often with
 factorization assumptions~\cite{desh,chau,xing,dean,fl,dav,kpsvv,kps,du}, has
 led to a number of these predictions, and the experimental sensitivity has
 now become sufficient to allow us to begin to test the correctness of the
 underlying assumptions~\cite{bigrare,kpi,etaetapr}.

 In this letter, we describe searches for $B$-meson decays to exclusive final
 states that include an $\omega$ or $\phi$ meson and one other low-mass
 charmless meson. Some decays to final states with a $\phi$ are of particular
 interest because they are dominated by penguin amplitudes, and receive no
 contribution from tree-level amplitudes (see Figure~\ref{f:diag}), while
 others, such as $B^+ \ra \phi \pi^+$, receive no contribution from penguin or
 tree amplitudes and only proceed through higher-order diagrams.

 The results presented here are based on data collected with the CLEO II
 detector~\cite{CLEOdet} at the Cornell Electron Storage Ring (CESR). The data
 sample corresponds to an integrated luminosity of 3.11~fb$^{-1}$ for the
 reaction $e^+ e^- \rightarrow \Upsilon(4S) \rightarrow B\overline{B}$, which
 in turn corresponds to $3.3 \times 10^6$ $B\overline{B}$ pairs. To study
 background from continuum processes, we also collected 1.61~fb$^{-1}$ of data
 at a center-of-mass energy below the threshold for $B\overline{B}$
 production.

 The final states of the decays under study are reconstructed by combining
 detected photons and charged pions and kaons. The $\omega$ and $\phi$ mesons
 are identified via the decay modes $\omega \ra \pi^+ \pi^- \pi^0$ and $\phi
 \ra K^+ K^-$, respectively. The detector elements most important for the
 analyses presented here are the tracking system, which consists of 67
 concentric drift chamber layers, and the high-resolution electromagnetic
 calorimeter, made of 7800 CsI(Tl) crystals.

 Reconstructed charged tracks are required to pass quality cuts based on their
 track fit residuals and impact parameter. The specific ionization ($dE/dx$)
 measured in the drift layers is used to distinguish kaons from
 pions. Expressed as the number of standard deviations from the expected
 value, $S_i$($i = \pi,K$), it is required to satisfy $|S_i| < 3.0$.
 Photons are defined as isolated showers, not matched to any charged tracks,
 with a lateral shape consistent with that of photons, and with a
 measured energy of at least 30~(50)~MeV in the calorimeter region
 $|\cos{\theta}| < 0.71 (\geq 0.71)$, where $\theta$ is the polar angle.

 Pairs of photons (charged pions) are used to reconstruct $\pi^0$'s and
 $\eta$'s ($K^0$'s). The momentum of the pair is obtained with a kinematic fit
 of the decay particle momenta with the meson mass constrained to its
 nominal value. To reduce combinatoric background, we reject very asymmetric
 $\pi^0$ and $\eta$ decays by requiring that the rest frame angle $\theta^*$
 between the direction of the meson and the direction of the photons satisfy
 $|\cos{\theta^*}| < 0.97$, and require that the momentum of charged tracks
 and photon pairs be greater than 100~MeV/$c$.

 The primary means of identification of $B$ meson candidates is through their
 measured mass and energy. The quantity $\Delta E$ is defined as $\Delta E
 \equiv E_1 + E_2 - E_b$, where $E_1$ and $E_2$ are the energies of the two
 daughter particles of the $B$ and $E_b$ is the beam energy. The
 beam-constrained mass of the candidate is defined as $M \equiv \sqrt{E_b^2 -
 |{\bf p}|^2}$, where $\bf p$ is the measured momentum of the candidate. We use
 the beam energy instead of the measured energy of the $B$ candidate to
 improve the mass resolution by about one order of magnitude.

 The large background from continuum quark--antiquark ($q\bar q$) production
 can be reduced with event shape cuts. Because $B$ mesons are
 produced almost at rest, the decay products of the $B\bar B$ pair tend to be
 isotropically distributed, while particles from $q\bar q$ production have a
 more jet-like distribution. The angle $\theta_{T}$ between the thrust axis of
 the charged particles and photons forming the candidate $B$ and the thrust
 axis of the remainder of the event is required to satisfy $|\cos{\theta_{T}}|
 < 0.9$. Continuum background is strongly peaked near 1.0 and signal is
 approximately flat for this quantity. We also form a Fisher discriminant
 (${\cal F}$)~\cite{bigrare} with the momentum scalar sum of
 charged particles and photons in nine cones of increasing polar angle around
 the thrust axis of the candidate and the angles of the thrust axis of the
 candidate and $\bf p$ with respect to the beam axis.

 The specific final states investigated are identified via the reconstructed
 invariant masses of the $B$ daughter resonances. For final states with a
 pseudoscalar meson, and for the secondary decay $\eta' \ra \rho \gamma$,
 further separation of signal events from combinatoric background is obtained
 through the use of the defined angular helicity state of the $\phi$,
 $\omega$, or $\rho$. The observable $\cal H$ is the cosine of the angle
 between the flight direction of the vector meson and the daughter decay
 direction (normal to the decay plane for the $\omega$), boosted to the
 meson's rest frame. For the final states $\omega K^{*+}$ and $\omega \rho^+$,
 the $\pi^0$ from $K^{*+}$ or $\rho^+$ decay defines the daughter direction.
 In this case we require ${\cal H}<0.5$ to reduce the large combinatoric
 background from soft $\pi^0$'s. Since the distribution of $\cal H$ is not
 known for these vector-vector final states we assume the worst case (${\cal
 H}^2$) when computing the efficiency.

 Signal event yields for each mode are obtained with unbinned multi-variable
 maximum likelihood fits. We also performed event counting analyses that
 applied tight constraints on all variables described above. Results for the
 latter are consistent with the ones presented below.

 For $N$ input events and $p$ input variables, the likelihood is defined as
\begin{eqnarray*}
 {\cal L} & = & e^{-(N_S+N_B)} \prod_{i=1}^N \{N_{S} \prod_{j=1}^p
   {\cal P}_{S_{ij}}(f_{1j},...,f_{mj};x_{ij}) + \\
   & & N_B \prod_{j=1}^p {\cal P}_{B_{ij}}(g_{1j},...,g_{nj};x_{ij})\} ,
\end{eqnarray*}
 where ${\cal P}_{S_{ij}}$ and ${\cal P}_{B_{ij}}$ are the probabilities for
 event $i$ to be signal and continuum background for variable $x_{ij}$,
 respectively. The probabilities are also a function of the parameters $f$ and
 $g$ used to describe the signal and background shapes for each variable. The
 number of parameters required varies depending on the input variable. The
 variables used are $\Delta E$, $M$, ${\cal F}$, resonance masses, and ${\cal
 H}$ as appropriate. For pairs of final states differentiated only by the
 identity of a single charged pion or kaon, we also use $S_i$ for that track
 and fit both modes simultaneously. $N_{S}$ and $N_B$, the free parameters of
 the fit, are the number of signal and continuum background events in the
 fitted sample, respectively. We verified that background from other $B$ decay
 modes is small for all channels investigated and did not require inclusion in
 the fit. Correlations between input variables were found to be negligible,
 except between the invariant masses of a parent resonance and its daughter,
 which the likelihood function takes into account.

 For each decay mode investigated, the signal probability distribution
 functions (PDFs) for the input variables are determined with fits to Monte
 Carlo event samples generated with a GEANT~\cite{geant} based simulation of
 the CLEO detector response. The parameters of the background PDFs are
 determined with similar fits to a sideband region of data defined by $|\Delta
 E| < 0.2$~GeV and $5.2 < M < 5.27$~GeV/c$^2$. The data samples collected on
 and below the $\Upsilon(4S)$ resonance are used. The signal shapes used are
 Gaussian, double Gaussian, and Breit-Wigner as appropriate for $\Delta E$ and
 mass peaks. For background, resonance masses are fit to the sum of a smooth
 polynomial and the signal shape, to account for the component of real
 resonance as well as the combinatoric background. For $\Delta E$ and $M$
 background we use a first-degree polynomial and the empirical shape $f(z)
 \propto M\sqrt{1-z^2}\exp{(-\xi(1-z^2))}$, where $z\equiv M/E_b$ and $\xi$ is
 a parameter to be fit, respectively. Finally, for ${\cal F}$, $S_K$, and
 $S_{\pi}$, we use bifurcated Gaussians for both signal and background.

 Sideband regions for each input variable are included in the likelihood
 fit. The number of events input to the fit varies from 70 to $\sim 12000$,
 depending on the final state. Table~\ref{omindivtab}~\cite{conj} gives the
 results for each mode investigated. The final state \omegah\ represents the
 sum of the \omegak\ and \omegapi\ states ($h^+ \equiv K^+$ or $\pi^+$). Shown
 are the signal event yield, the efficiency, the product of the efficiency and
 the relevant branching fractions of particles in the final state, and the
 branching fraction for each mode, given as a central value with statistical
 and systematic error, or as a 90\% confidence level upper limit. The one
 standard deviation ($\sigma$) statistical error is determined by finding the
 values where the quantity $\chi^2 = -2\ln({\cal L}/{\cal L}_{\rm max})$,
 where ${\cal L}_{\rm max}$ is the point of maximum likelihood, changes by one
 unit.

\begin{table}[htbp]
\caption{Measurement results.  Columns list the final states (with
secondary decay modes as subscripts), event yield from the fit,
reconstruction efficiency $\epsilon$, total efficiency including secondary
branching fractions ${\cal B}_s$, and the resulting $B$ decay branching
fraction ${\cal B}$.  }
\def\notext{ & & & & \cr}
\begin{center}
\begin{tabular}{lcrrc}
Final state & Yield(events) & $\epsilon$(\%) & $\epsilon\calB_s$(\%) &
\calB($10^{-5})$\cr
\sgline
\omegak       &$12.2^{+5.5}_{-4.5}$&28&25.1&$1.5^{+0.7}_{-0.6}\pm0.2$\cr
\omegakz      & $2.3^{+2.4}_{-1.5}$&15&      4.4      &$<5.7$  \cr
\omegapi      & $9.2^{+5.3}_{-4.3}$&29&     25.8      &$<2.3$  \cr
\omegah       & $21.4^{+6.5}_{-5.6}$&29&25.5&$2.5^{+0.8}_{-0.7}\pm0.3$\cr
\omegapiz     & $2.4^{+2.9}_{-1.8}$&24&     20.9      &$<1.4$  \cr
\omegaetaprd  & $0.1^{+1.9}_{-0.1}$&16&      2.4      &$<6.4$  \cr
\omegaetaprrg & $5.1^{+3.6}_{-2.7}$&16&      4.2      &$<9.2$  \cr
\omegaetagg   & $0.0^{+1.5}_{-0.0}$&24&      8.5      &$<2.0$  \cr
\omegaetathrp & $0.0^{+0.5}_{-0.0}$&15&      3.2      &$<2.8$  \cr
\omegakstpd   & $1.1^{+2.6}_{-1.1}$& 7&      2.0      &$<12.9$  \cr
\omegakstpkz  & $4.5^{+3.6}_{-2.8}$&16&      3.2      &$<10.9$ \cr
\omegakstzd   & $2.1^{+3.6}_{-2.1}$&22&     13.1      &$<2.3$  \cr
\omegarhop    & $2.5^{+4.4}_{-2.5}$& 8&      6.8      &$<6.1$  \cr
\omegarhoz    & $0.0^{+1.7}_{-0.0}$&24&     21.1      &$<1.1$  \cr
\omegaomega   & $0.3^{+2.6}_{-0.3}$&15&     11.9      &$<1.9$  \cr
\phik         & $0.0^{+0.8}_{-0.0}$&47&     23.1      &$<0.5$  \cr
\phikz        & $1.9^{+2.0}_{-1.2}$&32&      5.3      &$<3.1$  \cr
\phipi        & $0.0^{+0.9}_{-0.0}$&49&     24.0      &$<0.5$  \cr
\phipiz       & $0.0^{+0.6}_{-0.0}$&31&     15.1      &$<0.5$  \cr
\phietaprd    & $0.0^{+0.5}_{-0.0}$&26&      2.2      &$<3.5$  \cr
\phietaprrg   & $2.7^{+3.1}_{-2.1}$&30&      4.4      &$<6.3$  \cr
\phietagg     & $0.0^{+0.6}_{-0.0}$&39&      7.5      &$<1.3$  \cr
\phietathrp   & $0.0^{+0.5}_{-0.0}$&24&      2.7      &$<2.9$  \cr
\phikstpd     & $2.6^{+3.3}_{-2.4}$&26&      4.4      &$<5.6$  \cr
\phikstpkz    & $1.7^{+2.0}_{-1.1}$&29&      3.4      &$<5.3$  \cr
\phikstzd     & $3.2^{+3.2}_{-2.1}$&39&     12.7      &$<2.2$  \cr
\phikstzkz    & $0.0^{+1.9}_{-0.0}$&18&      1.0      &$<8.0$  \cr
\phirhop      & $0.0^{+2.3}_{-0.0}$&34&     16.7      &$<1.6$  \cr
\phirhoz      & $0.8^{+4.4}_{-0.8}$&41&     20.0      &$<1.3$  \cr
\phiomega     & $0.8^{+2.5}_{-0.8}$&23&     10.2      &$<2.1$  \cr
\phiphi       & $0.4^{+1.4}_{-0.4}$&40&      9.7      &$<1.2$  \cr
\end{tabular}
\end{center}
\label{omindivtab}
\end{table}

 Systematic errors are separated into two major components. The first is
 systematic errors in the PDFs, which are determined with a Monte Carlo
 variation of the PDF parameters within their Gaussian uncertainty, taking
 into account correlations between parameters. The final likelihood function
 is the average of the likelihood functions for all variations. The second
 component is systematic errors associated with event selection and efficiency
 factors. For cases where we determine a branching fraction central value, the
 final systematic error is the quadrature sum of the two components. For upper
 limits, the likelihood function including systematic variations of the PDFs
 is integrated to find the value that corresponds to 90\% of the total
 area. The efficiency is reduced by one standard deviation of its systematic
 error when calculating the final upper limit.

 For final states which we detect in multiple secondary channels, we sum the
 value of $\chi^2$ as a function of the branching fraction and
 extract the final branching fraction or upper limit from the combined
 distribution. Table~\ref{omcombtab} shows the final results, as well as
 previously published theoretical estimates.

\begin{table}[htbp]
\caption{Combined results and expectations from theoretical models.}
\vspace{0.4cm}
\def\notext{ & & \cr}
\begin{center}
\begin{tabular}{lccl}
Decay mode  &    \calB($10^{-5})$     & Theory \calB\ ($10^{-5}$)&References\cr
\sgline
\Bomegak    &$1.5^{+0.7}_{-0.6}\pm0.2$&$0.1-0.7$&\cite{chau,dean,kps,du} \cr
\Bomegakz   &          $<5.7$         &$0.1-0.4$&\cite{chau,dean,du} \cr
\Bomegapi   &          $<2.3$         &$0.1-0.7$&\cite{chau,dean,kps,du} \cr
\Bomegah    &$2.5^{+0.8}_{-0.7}\pm0.3$& -       & - \cr
\Bomegapiz  &          $<1.4$         &$0.01-1.2$&\cite{chau,dean,du} \cr
\Bomegaetapr&          $<6.0$         &$0.3-1.7$&\cite{chau,du} \cr
\Bomegaeta  &          $<1.2$         &$0.1-0.5$&\cite{chau,du} \cr
\Bomegakstp &          $<8.7$         &$0.04-1.5$&\cite{chau,dean,kpsvv} \cr
\Bomegakstz &          $<2.3$         &$0.2-0.8$&\cite{chau,dean} \cr
\Bomegarhop &          $<6.1$         &$1.0-2.5$&\cite{chau,dean,kpsvv} \cr
\Bomegarhoz &          $<1.1$         &$0.04$   &\cite{chau} \cr
\Bomegaomega&          $<1.9$         &$0.04-0.3$&\cite{chau,dean} \cr
\Bphik      &          $<0.5$
&$0.07-1.6$&\cite{desh,chau,dean,fl,dav,kps,du}\cr 
\Bphikz     &          $<3.1$
&$0.07-1.3$&\cite{desh,chau,dean,fl,dav,du}\cr 
\Bphipi     &          $<0.5$         & $<<0.1$ &\cite{xing,dean,fl,kps,du}\cr
\Bphipiz    &          $<0.5$         & $<<0.1$ &\cite{xing,dean,fl,du}\cr
\Bphietapr  &          $<3.1$         & $<<0.1$ &\cite{xing,du}  \cr
\Bphieta    &          $<0.9$         & $<<0.1$ &\cite{xing,dean,du}  \cr
\Bphikstp   &          $<4.1$
&$0.02-3.1$&\cite{desh,chau,dean,dav,kpsvv}\cr 
\Bphikstz   &          $<2.1$         &$0.02-3.1$&\cite{desh,chau,dean,dav} \cr
\Bphikst    &          $<2.2$         &$0.02-3.1$&\cite{desh,chau,dean,dav}\cr
\Bphirhop   &          $<1.6$         & $<<0.1$ &\cite{xing,dean,kpsvv} \cr
\Bphirhoz   &          $<1.3$         & $<<0.1$ &\cite{xing,dean} \cr
\Bphiomega  &          $<2.1$         & $<<0.1$ &\cite{xing,dean} \cr
\Bphiphi    &          $<1.2$         &none & \cr
\end{tabular}
\end{center}
\label{omcombtab}
\end{table}

\begin{figure}[htbp]
\psfile{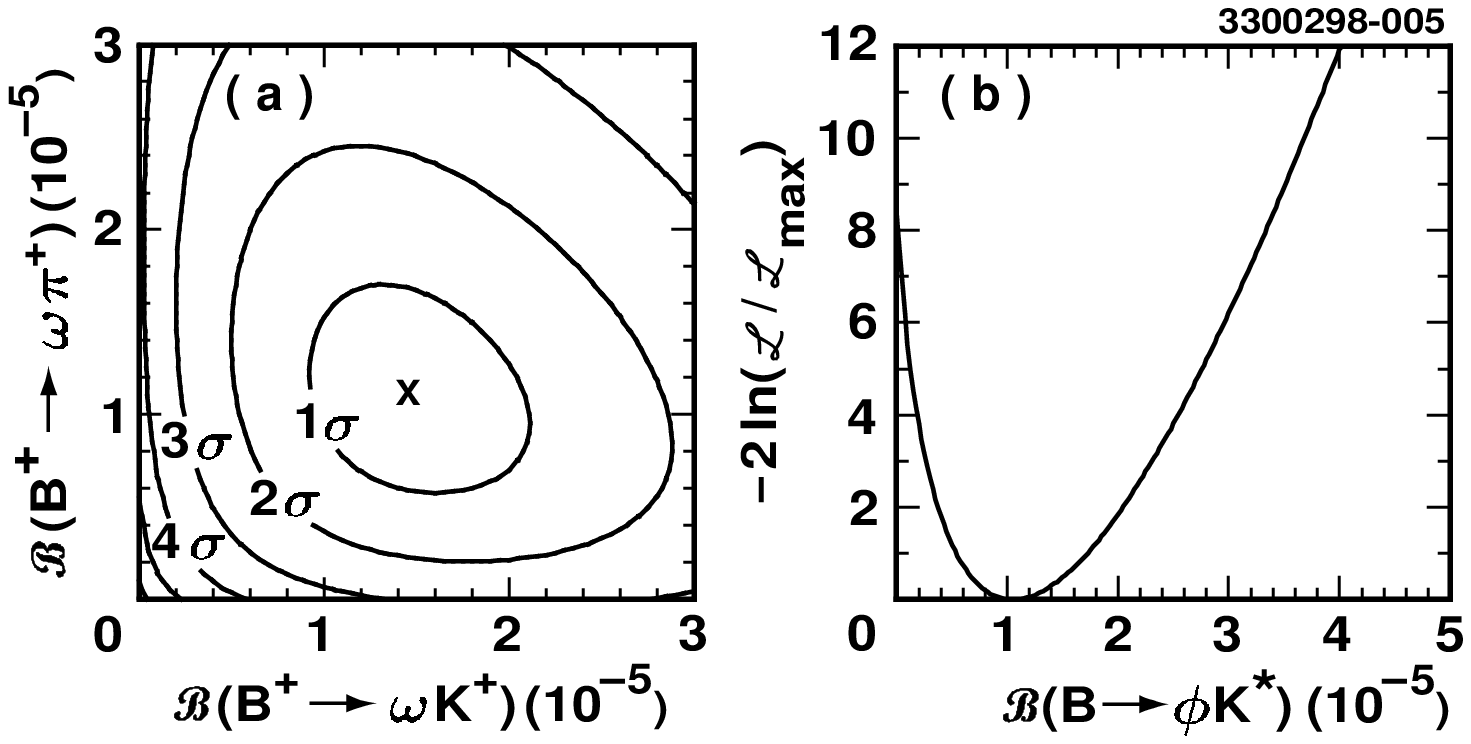}{0.9}
 \caption{\label{f:cont}%
(a) Likelihood function contours for $B^+\ra\omega h^+$; (b) The function
$-2\ln{{\cal L}/{\cal L}_{\rm max}}=\chi^2-\chi^2_{\rm min}$ for $B \ra \phi
K^*$.}
\end{figure}

\begin{figure}[htbp]
\psfile{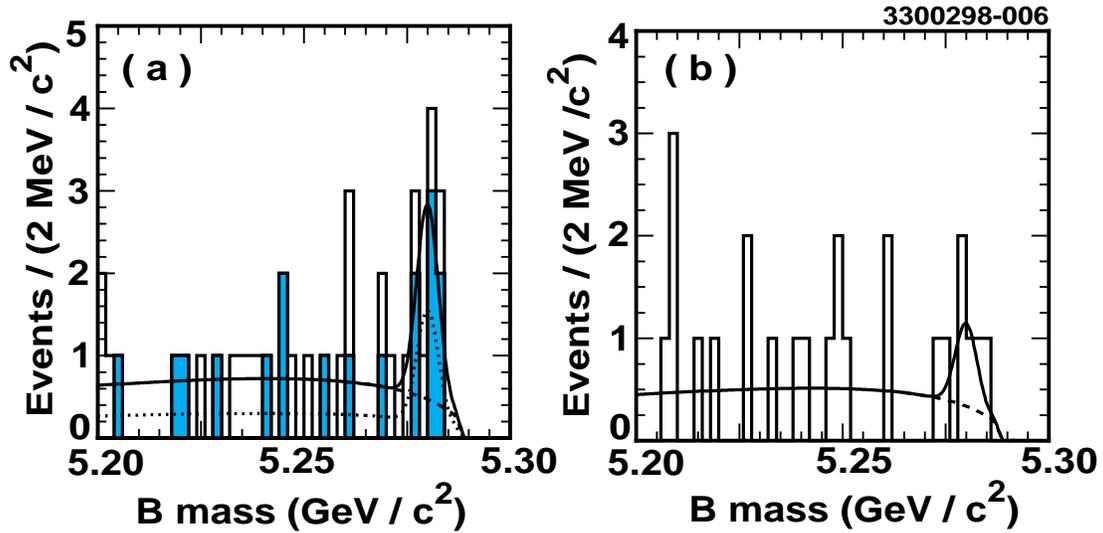}{0.9}
 \caption{\label{f:proj}%
 Projection onto the variable $M$ for (a) \Bomegak (shaded) and \Bomegapi
(open) and (b) \Bphikst. The solid line shows the result of the likelihood
fit, scaled to take into account the cuts applied to variables not
shown. The dashed line shows the background component, and in (a) the dotted
line shows the \Bomegak\ fit only.}
\end{figure}

 We find a significant signal for \Bomegak\ and measure the branching fraction
 ${\cal B}(B^+ \rightarrow \omega K^+) = (1.5^{+0.7}_{-0.6}\pm0.2)\times
 10^{-5}$, where the first error is statistical and the second systematic. We
 also find a signal for \Bomegah, with a branching fraction of ${\cal B}(B^+
 \rightarrow \omega h^+) = (2.5^{+0.8}_{-0.7}\pm0.3)\times 10^{-5}$. The
 significance for these signals is 3.9$\sigma$ for \Bomegak\ and 5.5$\sigma$
 for \Bomegah. We also find some evidence for the sum of the modes \Bphikstp\
 and \Bphikstz, with a significance of $2.9\sigma$. It is sensible to combine
 these modes since their decay rate is expected to be dominated by identical
 penguin amplitude contributions, except for different spectator quarks. The
 quoted significances include both statistical and systematic errors. If we
 interpret the observed $\phi K^*$ event yield as a signal, we obtain a
 branching fraction of ${\cal B}(B \rightarrow \phi K^*) =
 (1.1^{+0.6}_{-0.5}\pm0.2)\times 10^{-5}$.  Figure~\ref{f:cont} shows the
 likelihood functions for these modes. Figure~\ref{f:proj} shows the
 projection along the $M$ axis, with clear peaks at the $B$ meson mass.

 We also set lower limits on the branching fractions for
 \Bomegak\ and \Bomegah, which could have interesting theoretical
 implications~\cite{cheng,ciuchini,oh,dighe,ali}. We find ${\cal B}(B^+
\rightarrow \omega K^+) > 8.4 \times 10^{-6}$ and ${\cal B}(B^+ \rightarrow
\omega h^+) > 1.6 \times 10^{-5}$ at the 90\% confidence level. The latter
limit would imply that the parameter $\xi$ used in references \cite{oh} and
\cite{ali} is restricted to the region $\xi > 0.62$ and $\xi > 0.53$,
respectively. However, based on reference \cite{oh} our measurement of ${\cal
B}(B^+ \ra \phi K^+) < 0.5 \times 10^{-5}$ implies that $\xi < 0.27$ at the
90\% confidence level. Although there is still considerable uncertainty in the
theoretical model parameters, these limits illustrate the
difficulty in accounting for all our current results with a single
phenomenological parameter.

We thank A. Ali, H. Lipkin, J. Rosner, H.-Y. Cheng, and S. Oh for useful
discussions. We gratefully acknowledge the effort of the CESR staff in
providing us with excellent luminosity and running conditions.  This work was
supported by the National Science Foundation, the U.S. Department of Energy,
Research Corporation, the Natural Sciences and Engineering Research Council of
Canada, the A.P. Sloan Foundation, and the Swiss National Science Foundation.



\begin{references}
\bibitem{conj} Charge conjugate modes are implied throughout this paper.
\end{references}
\end{document}